\def\draftversion{true}
\RequirePackage{ifthen}
\ifthenelse{\equal{\draftversion}{true}}{

  \documentclass[aps,prb,10pt,twocolumn,amsmath,amssymb,
                 superscriptaddress,citeautoscript]{revtex4-2}
}{
  \documentclass[aps,prb,10pt,twocolumn,amsmath,amssymb,longbibliography,
                 superscriptaddress,citeautoscript]{revtex4-1}
}

\usepackage{graphicx}
\usepackage[usenames,dvipsnames]{color} 
\usepackage{bm} 
\usepackage{soul}

\usepackage{xr}
\usepackage{cleveref}
\ifthenelse{\equal{\draftversion}{true}}{
  \marginparwidth 2.7in
  \marginparsep 0.5in
  \newcounter{comm} 
  \def\commnext{\stepcounter{comm}}
  \def\commtext{{\bf\color{blue}[\arabic{comm}]}}
  \def\commmar{{\bf\color{blue}[\arabic{comm}]}}
  \def\dvm#1{\commnext\marginpar{\small DV\commmar: #1}\commtext}
  \def\sbm#1{\commnext\marginpar{\small SB\commmar: #1}\commtext}
  \def\mlab#1{\marginpar{\small\bf #1}}
  
}{
  \def\dvm#1{}
  \def\sbm#1{}
  \def\mlab#1{}
  
}
\newcommand{\beq}{\begin{equation}}
\newcommand{\eeq}{\end{equation}}
\newcommand{\bea}{\begin{eqnarray}}
\newcommand{\eea}{\end{eqnarray}}

\newcommand{\eqlab}[1]{\label{eq:#1}}
\newcommand{\figlab}[1]{\label{fig:#1}}
\def\z2{$\mathbb{Z}_2$}

\def\M{{\bf M}}

\def\H{{\bf H}}

\begin{document}


\title{A large anomalous Hall effect and Weyl nodes in bulk FeNi$_{3}$: a density functional theory study}

\author{Shivani Thakur}
\affiliation{Department of Physics and Materials Science, Jaypee University of Information Technology, Waknaghat, Solan, Himachal Pradesh 173234, India}

\author{Santu Baidya}
\email{santubaidya2009@gmail.com}
\affiliation{Department of Physics and Materials Science, Jaypee University of Information Technology, Waknaghat, Solan, Himachal Pradesh 173234, India}

\date{\today}
\begin{abstract}
	In this work, we report the study of electronic structure, magnetism and the existence of Weyl nodes in a pristine bulk FeNi$_{3}$, a member of Fe-Ni inver alloy compounds, known as good metal catalysts with high activity and stability for water splitting for a very long time. Our observation of Weyl points in the bulk FeNi$_{3}$ may lead to a new technology to design high-efficient topological catalysts. While the previous literatures \cite{PhysRev.97.304} mainly focused on the thermal and catalytic properties of FeNi$_{3}$ we report the interplay of Fe $d$-Ni $d$ hybridization and spin-orbit coupling give rise to the ferromagnetic Weyl nodes in the bulk FeNi$_{3}$. Our study shows that the ground state of the bulk FeNi$_{3}$ is a Weyl metal with a large number of Weyl nodes at the Fermi energy away from high-symmetry $k$-points. Furthermore, we predict a large intrinsic anomalous Hall conductivity of about $10000~S/m$ at the ground state. In addition, we show existence of Weyl nodes along the high symmetry $k$-points $~0.2eV$ above and $~0.05eV$ below self-consistent Fermi level that may be achieved either by electron or hole doping, or by external perturbation. In this article, FeNi$_{3}$ has been studied to explore this scenario using first-principles density functional theory and subsequent Wannier90 based tight-binding method. Furthermore, we report the existence of two types of Weyl cones, type-I and type-II, $~0.2eV$ above Fermi level. Our report provides a realistic material to further explore the intrinsic properties related to Weyl cones, and the spintronic applications.
\end{abstract}

\maketitle



%
%
%

\section{INTRODUCTION}
The magnetic Weyl metals/semimetals have attracted a lot of attentions in the past few years due to future applications in spintronic and quantum devices. Usually either by breaking inversion symmetry or time-reversal symmetry of a Dirac semimetal, one can obtain a Weyl semimetal \cite{PhysRevB.102.165115}. The magnetic Weyl metals/semimetals are very rare in nature to observe. There are very few materials such as Co$_{3}$Sn$_{2}$S$_{2}$ \cite{Okamura2020-fd}, NdAlSi \cite{Li2023-wi, PhysRevLett.128.176401}, etc. which were experimentally observed to have magnetic Weyl semimetal properties. With the purpose of searching new magnetic Weyl metal/semimetal along this line, we focus on a material FeNi$_{3}$ of the Fe-Ni invar alloy family, FeNi$_{3}$, Fe$_{3}$Ni, and FeNi, that attracted tremendous interests for their novel electronic, thermodynamic, and catalytic properties. The previous literatures \cite{PhysRev.97.304, PhysRevB.16.993, Cranshaw_1987} have reported magnetic, electronic ground state and thermodynamic properties of bulk FeNi$_{3}$. The most studied Pt-based catalysts remain irreplaceable and is still in use with doping \cite{YI2019113279}, surface modification \cite{ZHANG2020890}, and alloying. FeNi$_{3}$ is one such compound which was studied as an efficient catalyst with alloying, heterostructured \cite{Qayum}, and doping to use for water splitting \cite{Dong}. Other than catalysis, FeNi$_{3}$ can be used in advance sensitive applications. A high saturation magnetization and high permeability have made Fe-Ni alloys, regarded as traditional soft magnetic materials, a major focus. Under pressure, a study was carried out on the structural constants, elastic, electronic, and magnetic properties of three Fe-Ni binary metals (FeNi$_{3}$, FeNi, and Fe$_{3}$Ni) \cite{GEHRMANN}. While a ferromagnetic ordering was confirmed by both the experiment and the theoretical study in the previous literatures, the magnetic exchange interactions and the effect of spin-orbit coupling in the electronic structure of FeNi$_{3}$ is not explored extensively.\\
Furthermore, the presence of topologically non-trivial electronic bands near Fermi level may lead to interesting application such as topological catalysts \cite{Meinzer2017-me}. Observing such property in a well-known material such as FeNi$_{3}$ would open a new door to future applications. In the present manuscript, the FeNi$_{3}$ is studied to report interesting Weyl metallic properties previously unexplored as per our literature survey. The first-principles density functional theory (DFT) and Wannier90 based tight-binding method is used to study electronic structure, and topological Weyl metallic phase of FeNi$_{3}$.

In this paper we focus on the magnetic interactions and electronic topological properties of the ground state of FeNi$_{3}$ and anomalous Hall conductivtiy.  Our work is motivated by recent proposals that such catalysts with topologically nontrivial bands could have applications like topological catalysts and spintronic devices.

\section {Ground State Electronic Structure of FeNi$_{3}$}
\subsection{First-principles Calculation}
The invar-alloy FeNi$_{3}$ was studied in the present manuscript due to the presence of time-reversal symmetry from partially filled $d$-orbitals of Fe and Ni atoms. The fig. \ref{fig:dft_band}(a) shows the primitive unit cell of FeNi$_{3}$ with space group $Pm\bar{3}m$ cubic lattice. The Fe atoms (yellow balls) takes the corner positions of the primitive unit cell while the face centered positions are occupied by Ni atoms (grey balls). The FeNi$_3$ has an analogous structure to the intermetallic material CrPt$_3$ \cite{OPPENEER2002371,SUHARYADI2016186} which was reported as a topological metal\cite{Markou2021}. CrPt$_3$ is a ferromagnetic metal having ground state band structure originated from $d-d$ hybridization as reported in previous literature\cite{Markou2021}. Similar to CrPt$_3$, FeNi$_3$ is highly symmetric. However, when ferromagnetic ordering is considered, the symmetry reduces to a tetragonal magnetic symmetry $P4mm^\prime m^\prime$. When ferromagnetic ordering is considered along the crystallographic $[001]$-direction ($c-$axis), Ni-ion's symmetry on the top/bottom faces (Ni$_{1c}$) differentiates from side faces (Ni$_{2e}$). The PBE+SOC density of states (DOS) projected onto $d$-orbital of Ni$_{1c}$ and Ni$_{2e}$ shown in the Fig. \ref{fig:pdos} highlights the differentiation between two inequivalent Ni sites. The optimized atomic positions along with the Wyckoff positions are tabulated for the bulk FeNi$_3$ having cubic lattice constant $a=3.4079\AA$ in the Table \ref{tab:wyckoff}.
\begin{figure}
\includegraphics[scale=0.36]{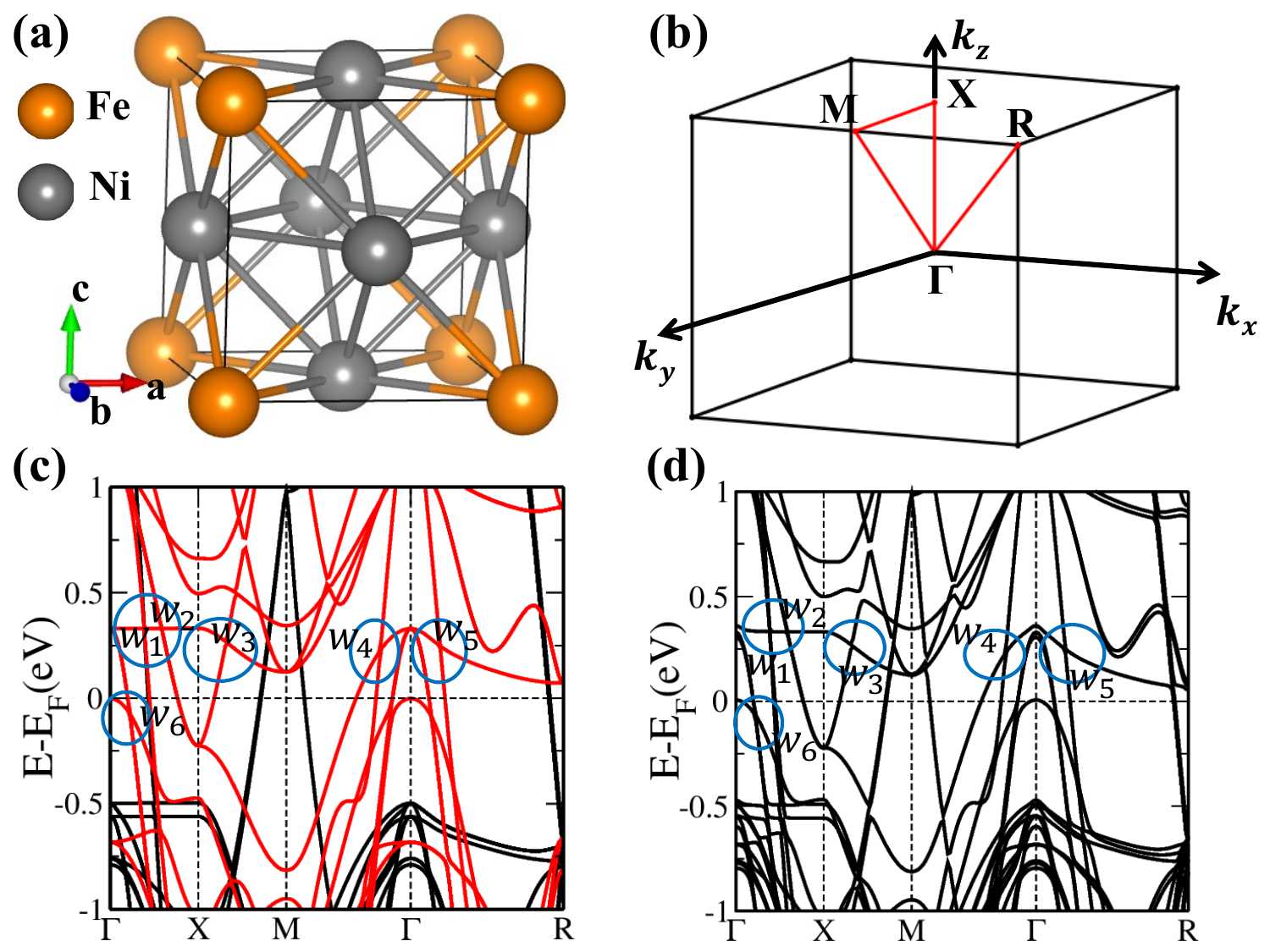}
	\caption{(a) The primitive unit cell of FeNi$_{3}$ with space group $Pm\bar{3}m$. (b) The Brillouin zone of the primitive unit cell with high-symmetry kpoints showing kpaths along which band structure is plotted. (c) The PBE band structure under FM ordering of Fe and Ni spins with linear crossings (blue circled). (d) PBE+SOC band structure with linear crossings (blue circled). A few crosing points (encircled) {\it{$w_{1}$}}, {\it{$w_{2}$}}, {\it{$w_3$}}, {\it{$w_4$}}, {\it{$w_5$}}, and {\it{$w_6$}} above and below Fermi level along the high-symmetry lines are highlighted.}
\figlab{dft_band}
\end{figure}

To perform DFT calculations, we used the plane-wave based program of Quantum ESPRESSO \cite{Giannozzi_2009} and used the projector augmented wave (PAW) pseudopotentials obtained from pslibrary \cite{DALCORSO2014337}. The valence configuration of Fe is (3d)$^6$(4s)$^2$ and Ni is (3d)$^8$(4s)$^2$ . The effects of exchange and correlation are treated within Perdew–Burke–Ernzerhof (PBE)\cite{PhysRevLett}. The effect of spin orbit coupling is treated within the scheme described in Refs. \cite{PhysRevB.71.115106}. The cutoff energy for the wavefunction is set to be $90$ Ryd with spin-orbit coupling (SOC), respectively. The size of the cutoff energy for the charge density is increased to ten fold that for the wavefunction. The self-consistent field (scf) calculations were performed by using $8\times8\times8$ {\it k}-Monkhorst-Pack grid. The total energy is converged within  $1.90\times 10^{-10}$Ryd in the scf calculation. The scf band structure calculations were crosschecked with OpenMx software package\cite{PhysRevB.67.155108,PhysRevB.69.195113} based on density functional theory (DFT), norm-conserving pseudopotentials, and pseudo-atomic localized basis functions.

\begin{table}[]
\centering
\caption{The atomic positions and Wyckoff positions for bulk FeNi$_{3}$ of space group Pm$\bar 3$m.}
\label{tab:wyckoff}
\begin{tabular}{c|c|c|c|c} 
\hline
 Atoms & $x$ & $y$ & $z$ & Wyckoff \\ \hline
 Fe& 0 & 0 & 0 & 1a \\
 Ni& 0.5 & 0.5 & 0 & 3c \\
 Ni& 0.5 & 0 & 0.5 & 3c \\
 Ni& 0 & 0.5 & 0.5 & 3c	\\\hline
\end{tabular}
\end{table}

\begin{figure}[ht]
\centering
\includegraphics[scale=0.33]{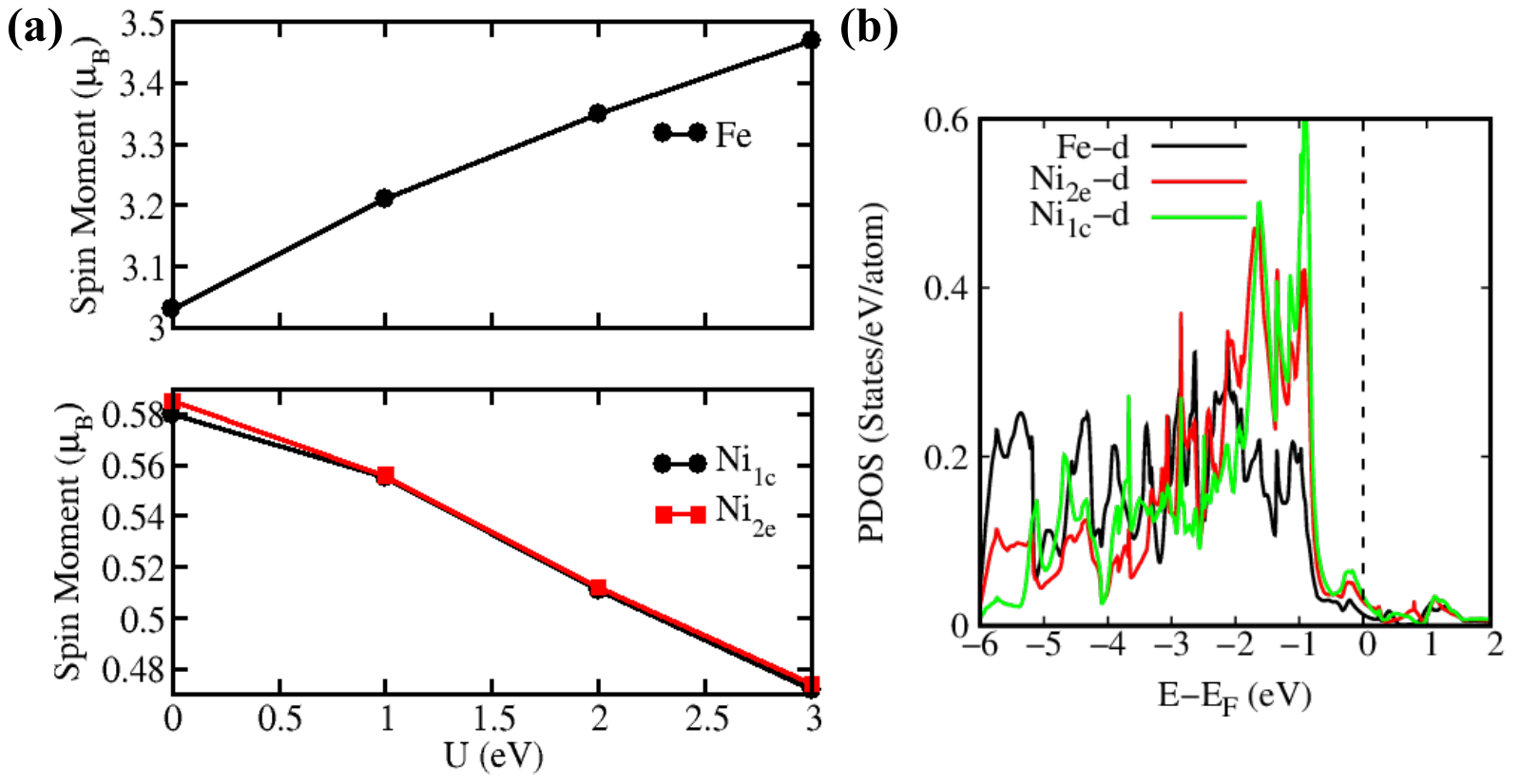}
	\caption{(a) Variation of spin magnetic moment of Fe and Ni ($1c$ and $2e$) sites under PBE+SOC approximation with Hubbard $U$ showing anisotropy in Ni sites even for finite values of Hubbard $U$. (b) PBE+SOC $d$-orbital projected density of states (PDOS) per atom shows finite density of states at the Fermi level. The distinct $d$-orbital projected DOS of the two Ni sites is visible.}
\figlab{pdos}
\end{figure}

First of all a density functional theory (DFT) calculation was carried out to get the ground state electronic band structure under PBE and PBE+SOC approximation using quantum espresso \cite{Giannozzi_2009} as shown in fig. \ref{fig:dft_band}(c) and fig. \ref{fig:dft_band}(d) respectively along a high-symmetry $k$-points  shown in red color lines in fig. \ref{fig:dft_band}(b) with full Brillouin Zone of FeNi$_{3}$. The metallic band structure clearly shows the covalent character of the orbitals at the self-consistent field (scf) Fermi level. The $d$-orbital projected density of states of Fe shows larger spin splitting compared to that of $d$-orbital projected density of states of Ni-atoms ($1c$ and $2e$) under PBE approximation (without SOC). In the Fig. \ref{fig:pdos}(b), the presence of finite density of states from both Fe and Ni $d$ oribitals at the scf Fermi level confirms metallic ground state of FeNi$_{3}$. The larger spin-splitting gives rise to finite spin moment at Fe $~3.03 \mu_B$, at Ni$_{1c}$ $~0.584 \mu_B$ and at Ni$_{2e}$ $~0.585 \mu_B$. The total spin moment is $~4.8 \mu_B$/(formula unit) under PBE+SOC approximation. The orbital moment for Fe is $~0.053 \mu_B$, Ni$_{1c}$ $~0.040 \mu_B$, and Ni$_{2e}$ $~0.038 \mu_B$. The variation of spin moment with Hubbard $U$ under PBE+U+SOC approximation is presented in the Fig. \ref{fig:pdos}a to highlight the effect of localization of atomic orbitals as both Fe and Ni have localized $3d$ orbitals. The ferromagnetic ordering arises from the strong Fe $d$-Ni $d$ orbital interactions due to the profound Fe-$d$ Ni-$d$ hybridization evident from fig. \ref{fig:pdos}(b). The strong $d-d$ hybridization and spin-orbit coupling led to a large magnetocrystalline anisotropy $1~meV/$unitcell with an easy axis oriented along [001] direction, which is surprisingly quite large for a bulk material. 

To get microscopic understanding of electronic properties the band structures correspdoning to the ground state are plotted in fig. \ref{fig:dft_band}(c,d). In the absence of spin-orbit coupling, there are multiple linear crossings above and below the scf Fermi level shown in the fig. \ref{fig:dft_band}(c) using blue circles, along with the high symmtery $k$-points in the Brillouin zone (BZ). In fig. \ref{fig:dft_band}(c) the black colour lines represent the up-spin channel bands and red colour lines represent down-spin channel bands. The addition of spin-orbit coupling under PBE+SOC approximation breaks the degeneracy of multiple band crossings leaving many crossings as shown by blue circles in the Fig. \ref{fig:dft_band}(d). At the scf Fermi level there are no linear band crossings visible along the high-symmetry directions. On the other hand, there are multiple linear band crossings $~0.2$ eV above and $~0.05$ eV below the scf Fermi level marked as $w_1$, $w_2$, $w_3$, $w_4$, $w_5$, and $w_6$ in the Fig. \ref{fig:dft_band}(d) under PBE+SOC approximation. Although, by either electron doping or hole doping one can tune the Fermi level to the linear band crossing, the positions of the crossings at the scf Fermi level are more important for further investigations.

\begin{figure}[ht]
\centering
\includegraphics[scale=0.28]{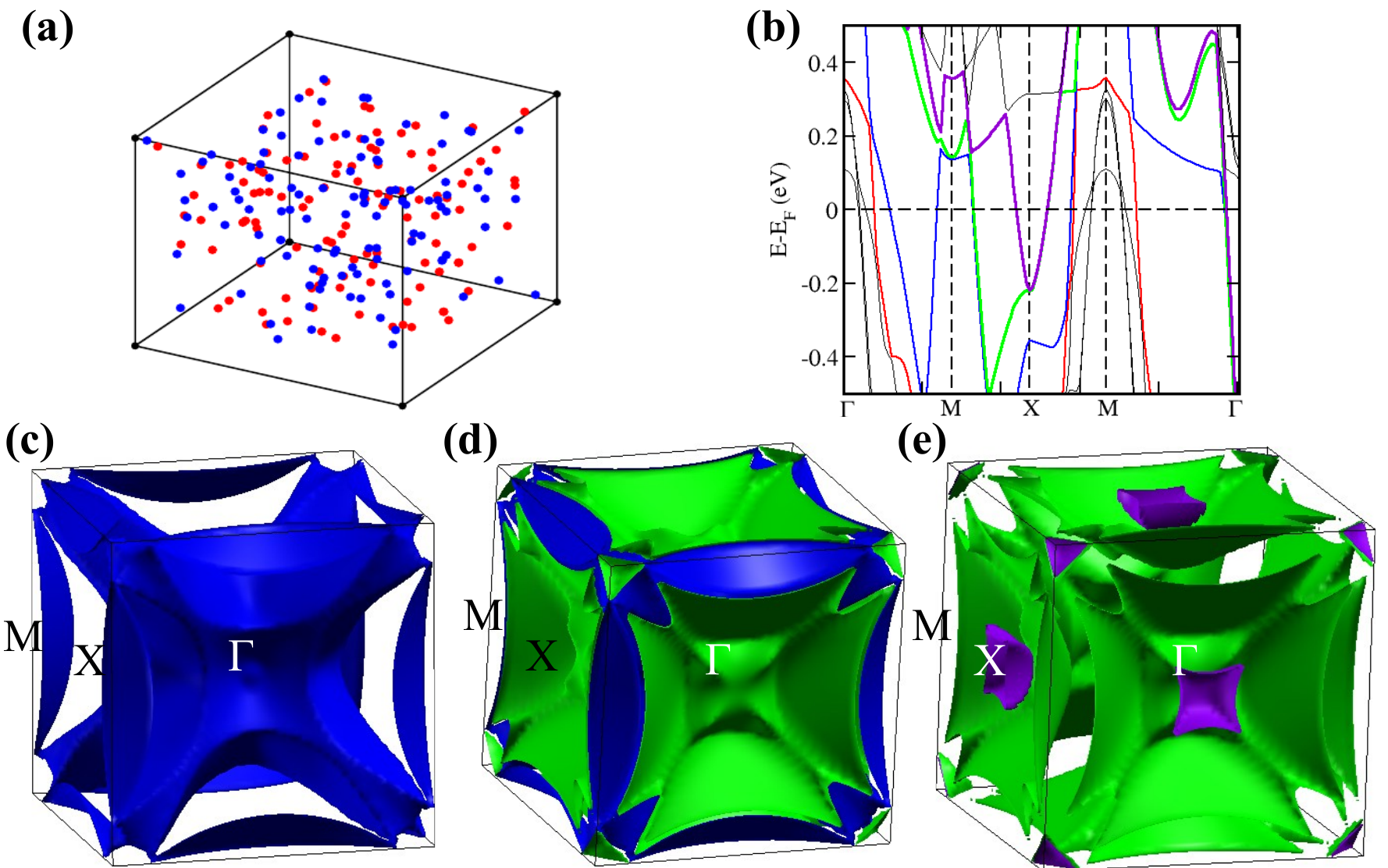}

\caption{(a) The Weyl map at the scf Fermi energy $E_F$ is shown in the reciprocal space that shows a large number of nodes away from high-symmetry planes, making it a Weyl metal with complicated linear band crossings. (b) The PBE+SOC band structure with few colored bands crossing the Fermi energy $E_F$, for which the Fermi surfaces are shown. (c) The Fermi surfaces of band $1$ (red) and band $2$ (blue). (d) The Fermi surfaces of band $3$ (blue) and band $4$ (green).(e) The Fermi surfaces of band $4$ (green) and band $5$ (purple).}

\figlab{weylmap}
\end{figure}

\subsection{Positions of Weyl points} 
We used the Wannier90 \cite{MOSTOFI20142309} package to acurately interpolate the band structure obtained from DFT and then used WannierTools \cite{WU2017} to calculate the topological charge or chiralities ($\chi$) corresponding to the linear band crossing points at the scf Fermi level. It gave a total of $112$ pairs of Weyl nodes with nonzero chiralities ($\chi$) whose locations are shown in the Fig. \ref{fig:weylmap}(a). The colours of Weyl nodes reflect their chiralities, with red and blue dots representing nodes with chiralities $+1$ and $-1$, respectively. Because the magnetic point group has eight symmetry elements total number of Weyl nodes should be in multiple of eight. The list of Weyl nodes with chiralities ($\chi$) is tabulated in the supplementary document. From the Weyl map it is evident that similar to CrPt$_3$ \cite{Markou2021} the FeNi$_3$ has complicated band crossings at the scf Fermi level making it a robust metal with a topological electronic structure. In fig. \ref{fig:weylmap}(b), the PBE+SOC band structure with few coloured band crossings at the scf Fermi level are shown. The bulk Fermi surfaces corresponding to the two bands red and blue is shown in Fig. \ref{fig:weylmap}(c). The bulk Fermi surfaces corresponding to the two bands blue and green are shown in Fig. \ref{fig:weylmap}(d). The bulk Fermi surfaces corresponding to the two bands green and purple are shown in Fig. \ref{fig:weylmap}(e). The Fermi surfaces at the scf Fermi energy shows the multiple band touching points inside the bulk Brillouin zone away from high-symmetry Brilluin zone planes.

In the Fig.\ref{fig:weyl-points}, a few high-symmetry linear band-touching points above and below the scf Fermi level are shown to show different types of Weyl cones in the bulk FeNi$_3$. In the Fig. \ref{fig:weyl-points}(a) the band touching point $w_1$ shows a type II Weyl cone. Fig. \ref{fig:weyl-points}(b-d) shows a type I Weyl cone. To further justify that these linear band touching points are not any avoided crossings, the irreducible representations of the crossing bands at the touching points $w_2$ (\ref{fig:weyl-points}(b)) and $w_3$ (\ref{fig:weyl-points}(c)) are shown. The irreducible representation of the two bands crossing the nodes $w_3$ and $w_4$ are marked as $G_3$ (red plus symbol) and $G_4$ (blue plus symbol) of the magnetic double point group $D_2$. It is evident that two crossing bands belong to different irreducible representations of $D_2$ magnetic point group, so there is no avoided crossing. Instead, these crossing points are Weyl nodes. Furthermore, the energy positions of the high-symmetry band-touching points are quite close to the scf Fermi level making them accessible either by electron or hole doping.

\begin{figure}[ht]
\centering
\includegraphics[scale=0.53]{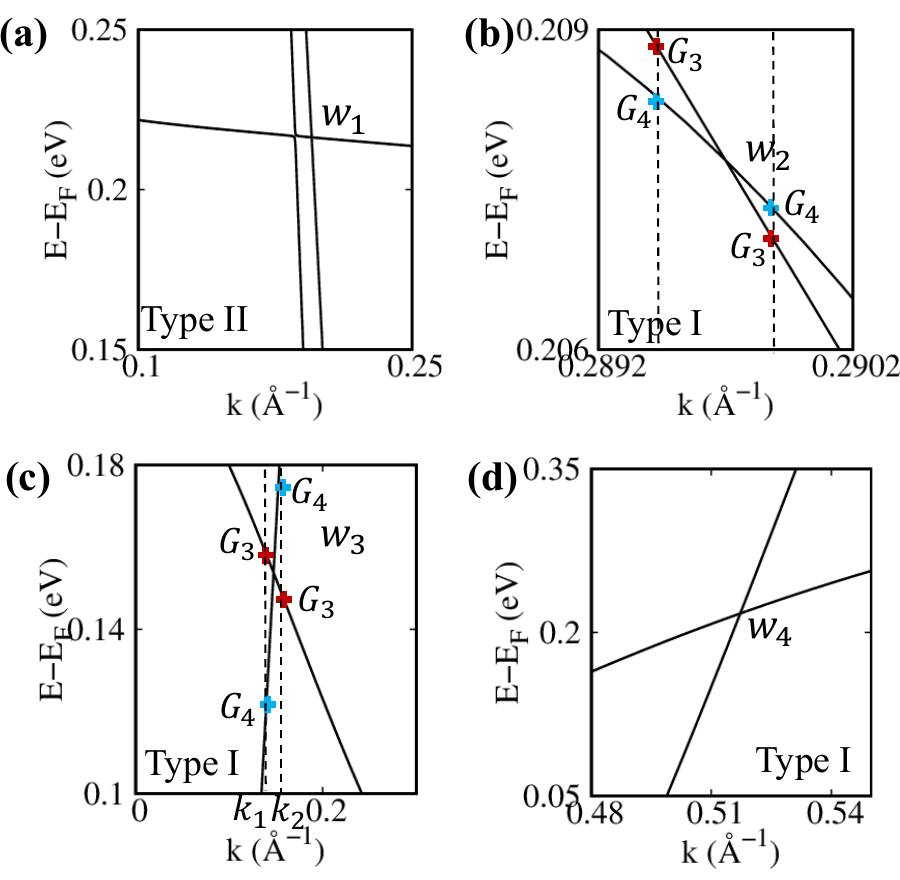}                                                                        

	\caption{PBE+SOC band structure shows (a) the position of Weyl point {\it{$w_1$}} at $k_{w1}$ at the $0.22$eV above scf $E_F$ along $\Gamma-X$ direction,(b) the position of Weyl point {\it{$w_2$}} at $k_{w2}$ at the $0.21$eV above scf $E_F$ along $\Gamma-X$ direction (c) the position of Weyl point {\it{$w_3$}} at $k_{w3}$ at $0.15$eV above scf $E_F$ along $X-M$ direction. The points with the plus symbols (red and blue color) represent the two $k-$points where the irreducible representations of the two bands crossing at the node $w_3$, are marked as $G_3$ and $G_4$ of magnetic double point group $D_2$. (d) {\it{$w_4$}}  at $k_{w4}$ at the $0.22$eV above scf $E_F$ along $M-\Gamma$ direction.}
\figlab{weyl-points}
\end{figure}

\begin{figure}                                                                                                \centering\includegraphics[scale=0.7]{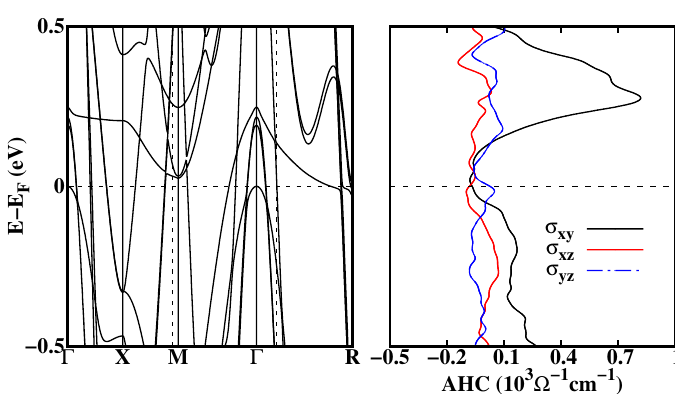}                                                                       \caption{The PBE+SOC band structure along with the variation of anomalous Hall conductivity matrix ($\sigma_{ij}$) with the variation of Fermi energy is shown.}                                                    \figlab{ahc}
\end{figure} 

\subsection{Anomalous Hall conductivity}
As FeNi$_3$ is ferromagnetic metal with the presence of Weyl nodes of nonzero chiralities at the scf Fermi level, we calculated the anomalous Hall conductivity ($\sigma_{ij}$;i,j=x,y or y,z or x,z) using Wannier90 and WannierTools package. Because the linear band crossings exist both above and below scf Fermi level, we varied the Fermi level to study the variation of anomalous Hall conductivity ($\sigma_{ij}$) across the Weyl nodes as shown in the Fig. \ref{fig:ahc} within the chosen energy manifold. A large anomalous Hall conductivity ($\sigma_{xy}$, black line in Fig. \ref{fig:ahc}) of $~10000$ (S/m) at the Fermi level was observed. As the Fermi level was varied above the scf Fermi level the anomalous Hall conducvity increases extremely high to $~70000$ (S/m) around $0.22$ eV. The anomalous Hall conducitivity ($\sigma_{ij}$;i,j=x,y or y,z or x,z) was calculated using the equation as implemented in WannierTools package

\begin{align}
	\sigma_{ij}& = \frac{e^2}{\hbar} \sum\limits_{BZ} \int \frac{\mathrm{d^3}k}{(2\pi)^3} \Omega_{ij}^n(\bm{k}) f{(E_{\bm{k}})}
\eqlab{ahceq}
\end{align}

Where $f{(E_{\bm{k}})}$ is the probability of occupation, and  $\Omega_{ij}^n(\bm{k})$ is the Berry curvature of  band, which is integrated over the whole Brillouin zone. The practical implication of the variation of Fermi level may be achieved by using a reasonable external pressure \cite{PANDYA2016287,Wang_2020} keeping symmetry of the materials unchanged. 

\section {Summary and conclusions}
With the purpose of studying consequence of ferromagnetic ordering on the electronic properties of bulk intermetallic Fe-Ni inver alloy such as FeNi$_{3}$ a detailed theoretical study on the electronic structure and topological properties of the bulk FeNi$_{3}$ is presented in this article. The ferromagnetic metallic ground state obtained in our theoretical calculation agrees very well with the previous literatures. Inspite of having localized $3d$ orbitals of Fe and Ni atoms, the bulk FeNi$_3$ shows highly covalent nature as evident from band analysis. The spin-orbit coupling give rise to topologically nontrivial characteristics via formation of Weyl nodes at the scf Fermi level. We predict the presence of multiple high-symmetry Weyl nodes in the bulk FeNi$_{3}$ which was previously unexplored. For the first time, we predict the presence of both tilted type-I and type-II Weyl cones in the Fe-Ni inver material FeNi$_{3}$, a well-known catalyst, about $~0.2$eV above the Fermi level along the high-symmetry direction in the Brillouin zone. The result of a large anomalous Hall conductivity at the scf Fermi level and an extremely large anomalous Hall conductivity $0.2$eV above the Fermi level should draw further attention of experimentalists and theoreticians for further exploration of this material. Our study may help in further studying Fe-Ni inver materials for understanding physics of topological catalysts and applications in superconductivity due to presence of  Weyl nodes.

\section{ASSOCIATED CONTENT}

{\bf Supporting Information} \\

Additional information on the list of Weyl nodes in the Brillouin zone is given in the supplementary information.

\bibliography{ref}
\end{document}



\title{Supplemental Information for "A large anomalous Hall effect and Weyl nodes in bulk FeNi$_{3}$: a density functional theory study"}

\author{Shivani Thakur}
\affiliation{Department of Physics and Materials Science, Jaypee University of Information Technology, Waknaghat, Solan, Himachal Pradesh 173234, India}

\author{Santu Baidya}
\email{santubaidya2009@gmail.com}
\affiliation{Department of Physics and Materials Science, Jaypee University of Information Technology, Waknaghat, Solan, Himachal Pradesh 173234, India}

\date{\today}

\maketitle



%
%
%

\section{LIST OF WEYL NODES}
The PBE+SOC band structure shows a total $112$ pairs of Weyl nodes away from high-symmetry planes of the Brillouin zone, at the scf Fermi level. The Chiralities ($\chi$) and the positions of the Weyl nodes at the scf Fermi level were obtained using WannierTools and Wannier90 packages. The list of Weyl nodes is tabulated below. 

\begin{table}[]
\centering
\caption{The list of positions of $28$ pairs of Weyl nodes with the Chiralities $+1$ and $-1$ at the scf Fermi level.}
\label{tab:weyllist1}
 {

\begin{tabular} {|c|c|c|c|c|c|c|c|}
\hline
k$_{x}$  & k$_{y}$ & k$_{z}$ & $\chi$ & k$_{x}$  & k$_{y}$ & k$_{z}$ & $\chi$  \\\hline
\hline
-0.47509 &  -0.46748 & 0.47908  & -1          &      -0.45032 & -0.44031 &  0.44351&  1 \\
 -0.38441 &  -0.40677 & -0.33078 & -1          &      0.31418 & -0.19210 &  0.32661&  1  \\
 -0.49331 & -0.24047   &-0.18655  & -1        &      0.49997 & -0.20027 & -0.23782&  1  \\
  0.46523 & -0.18396  & 0.03056  & -1          &    -0.49615 & -0.21031 & -0.15624 & 1 \\
  -0.44085 & -0.16423  & 0.23558 &  -1         &     0.47260  & -0.08904 & -0.02469 & 1\\
   -0.47866 & -0.24117  & 0.20466  &   -1       &   -0.48065 & -0.21473 & -0.00900& 1\\
   -0.49864 & -0.18024  & 0.04139  &   -1       &   0.47260 &  -0.08904 & -0.02469& 1 \\
   -0.49354 & -0.01765  & 0.18052  &    -1      &  -0.46828 & -0.17460  & 0.07175 & 1\\
    -0.46410 & -0.05796 & -0.04247 &     -1      &  -0.46367 & -0.13621  & 0.19896 &1 \\
     -0.49181 &  0.09716 &  0.01682 &    -1      & -0.45389 &  0.03152  & 0.11836 & 1 \\
     -0.49788 &  0.07814  & -0.20735 &    -1     & -0.49211 &   0.18132 &  0.04932 & 1 \\
      -0.49644 &  0.23161 & -0.20787  &    -1    &  -0.48840  & 0.18240 & -0.05024 & 1\\
       0.49784  & 0.23656  & 0.12665   &    -1   &  0.49535  &  0.23705 &  0.19106 & 1\\
      0.45549  & 0.43880  & -0.44750  &    -1    &  0.49888 &  0.22903  & 0.22119  &  1\\
      -0.23163  & 0.09361  & 0.19709  &      -1 &  -0.44480 & -0.16994 & -0.04429  & 1\\
        -0.21897 &  0.49767 & -0.23153 &      -1&  -0.39929 &  0.04028 & -0.13888  &1\\
         -0.45593 & -0.24005 &  0.00191 &    -1&  -0.40201 &  0.13840 &  -0.05091  & 1\\
         -0.46410 & -0.05796 & -0.04247 &    -1 & -0.44908  & 0.24080 & -0.02138  & 1\\
         -0.37921 &  0.03933  & 0.12502  &    -1&  -0.21768 &  0.49780  &  0.23505 & 1\\
         -0.43035 &  0.19086  & 0.04950  &     -1& -0.23482 & -0.45846  & 0.01732  & 1\\
          -0.31143 &  0.36419 &  0.10675  &    -1 & -0.23127 & -0.15432 & -0.19766  &1\\
          -0.29927  & 0.15549 & -0.11461  &     -1&  -0.17594 & -0.19501 &   0.18171 &1\\
           -0.28223  & 0.34409 &  0.42369  &   -1&  -0.19913 & -0.16846 &   0.35006 & 1\\
            -0.21231  & -0.18257&  -0.16726 &   -1& -0.23683  & 0.02123 & -0.45519 & 1\\
             -0.18108 & -0.40517 &  0.28732 &   -1& -0.35718  &  0.05416 &   0.11091 & 1\\
              -0.20604 & -0.14003 &  0.21307 &  -1&  -0.18334  & 0.14043 & -0.43887  & 1\\
               -0.34490 & -0.09605 &  0.04631 &  -1& -0.20703 &  0.43539 & -0.09181  & 1\\
               -0.25339  & 0.01333  & 0.43737  &  -1 &  -0.19942 & -0.48242 & -0.24182  & 1\\\hline
\end{tabular}
}
\end{table}

\begin{table}[]
\centering
\caption{The list of positions of $28$ pairs of Weyl nodes with the Chiralities $+1$ and $-1$ at the scf Fermi level.}
\label{tab:weyllist1}
 {
\begin{tabular} {|c|c|c|c|c|c|c|c|}
\hline
k$_{x}$  & k$_{y}$ & k$_{z}$ & $\chi$ & k$_{x}$  & k$_{y}$ & k$_{z}$ & $\chi$  \\\hline
	-0.28581 &  0.28320 & -0.46071 &  -1&  -0.18513 & -0.21676 &  0.49362  & 1\\
                 -0.19487  & 0.45818  & 0.10301 &   -1&  -0.21128 & -0.17548 & -0.41718 & 1\\
                 -0.20530 & -0.18200  & 0.46618 &    -1 &  -0.24562 & -0.17010  & 0.14246 & 1\\
                 -0.19423  & -0.19657  & 0.36716 &    -1&  -0.22575 &  -0.11526 &  0.49603& 1\\
                 -0.21766 & -0.00596  & -0.23328 &     -1&   -0.23613 & -0.00919 &  0.04840 & 1\\
                  -0.24083 & -0.00151 & -0.02712 &      -1&    -0.23232 &  0.01784 &  0.21082 & 1 \\
                  -0.20965  & 0.05845  & 0.49638  &      -1&  -0.19890 &  0.15398  & 0.33062&1\\
                  -0.20020   & 0.11112 & -0.42880 &       -1&  -0.21306 &  0.13187 &  0.48720 & 1\\
                   -0.14159  & 0.17937 & -0.27458  &      -1& -0.20993 &  0.22155 & -0.01511& 1\\
                    -0.22648  & 0.23732 & -0.48710 &        -1&  -0.14517 &  0.22289 & -0.22228& 1\\
                    -0.20742  & 0.20204  & 0.40864 &        -1&   -0.21425 &  0.24161  & 0.48323 & 1\\
                     -0.18656  & 0.43821 & -0.15656 &       -1&    -0.20013 &  0.44208  & 0.11208  & 1\\
                     -0.12442  & -0.46482 & -0.15865 &     -1&  -0.16374 & -0.47381 &  -0.19838 & 1\\
                     -0.10871  & -0.37742 &  0.08525  &    -1&   -0.09874 & -0.49760  & 0.20759 & 1\\
                      0.01721  & -0.45308  & 0.13497   &   -1&   -0.10432 & -0.39658 &  0.12752 & 1\\
                      -0.04744 & -0.22299 &  0.22043   &    -1&  -0.08941 & -0.11567 &  0.33647& 1\\
                      -0.15388 & -0.15250 &  0.43746   &     -1&  -0.12747 & -0.29554 & -0.39154& 1 \\
                       -0.11434 & -0.03911 & -0.43049  &      -1&   -0.16068 & -0.30767  & 0.09613 & 1\\
                       -0.20863 &  0.00877 &  0.09002  &      -1&    -0.02363 & -0.27942 &  0.00063 & 1\\
                       -0.07726  & 0.16321 & -0.47455  &       -1&   -0.10956 & -0.02343 & -0.44340 & 1\\
                       -0.06212  & 0.20427 & -0.49769  &       -1&   -0.14260 &  0.07342 & -0.32602 & 1\\
                        -0.15292  & 0.31672 & -0.09039  &       -1&   -0.06177 & -0.17209 &  0.13001 & 1\\
                         0.01085  & 0.42077 & -0.27278  &      -1 &   -0.11183 &  0.15473 & -0.43601 & 1\\
                         -0.09299 & -0.32707 & -0.36378 &   -1 &     -0.12884 &  0.14830 &  0.28785 & 1\\
                         -0.11988 & -0.23380 & -0.49084 &    -1&   -0.13178 &  0.28007 & -0.36827 & 1\\
                          -0.02729 & -0.20793 &  0.10257 &  -1&    -0.19524 &  0.46286 & -0.18624  &1\\
                           -0.02477 & -0.10565 &  0.19871 &  -1&  -0.13698 &  0.44522 &  0.00320  & 1\\
                           -0.00074 & -0.00066  & 0.02789 &  -1&   0.02936 & -0.42966 & -0.26426  & 1\\\hline
\end{tabular}
}
\end{table}


\begin{table}[]
\centering
\caption{The list of positions of $28$ pairs of Weyl nodes with the Chiralities $+1$ and $-1$ at the scf Fermi level.}
\label{tab:weyllist1}
 {
\begin{tabular} {|c|c|c|c|c|c|c|c|}
\hline
k$_{x}$  & k$_{y}$ & k$_{z}$ & $\chi$ & k$_{x}$  & k$_{y}$ & k$_{z}$ & $\chi$  \\\hline
0.00095 &  0.04102 & -0.24811 &  -1 &       -0.00615 & -0.22913 &  0.21132  & 1\\
-0.07848 &  0.13587 &  0.32266& -1 &   0.00599 & -0.23327 & -0.21064  & 1\\
-0.03686  & 0.17690 &  0.48922&  -1&   -0.00038 & -0.00096 & -0.02777  & 1\\
-0.00161 &  0.20192 & -0.23705&  -1&   -0.01462 & -0.04038  & 0.23868  & 1\\
0.00273  & 0.24693 & -0.03655&  -1&    -0.03822 &  0.01135  & 0.24993  & 1\\
-0.06870  & 0.25933  & 0.43693 &-1&   -0.02062 &  0.19735  & 0.24381  & 1\\
-0.03671 &  0.42980 & -0.13414 & -1& -0.01685 &  0.18543 &  0.43968  & 1 \\
0.01721 & -0.45308  & 0.13497  & -1&  0.01996 &  0.24122 & -0.00297 & 1\\
0.10145 &   0.44714 &  0.17263&  -1&   -0.06318 &  0.34989 & -0.12490 & 1\\
0.08483&  -0.31614 &  -0.15107 & -1 &   0.02095 &   0.24896 &  0.02378 & 1\\
0.10106 & -0.35914 &  0.07169 & -1 &    -0.01222 &  0.47324 &  0.21413& 1\\
0.08227 &  -0.11316 &  0.31496 & -1&   0.06061 & -0.17064 &  -0.46567 & 1\\
0.03036 &  -0.07944 & -0.21993 & -1 &  0.08801&   0.04891 &  0.42234  & 1\\
 0.04256 &  -0.00121 &  -0.24790 & -1 &  0.07387 & -0.09027 &   0.32537& 1\\
0.05718  & 0.12021 &  0.33076 & -1 &    0.08090 &  0.03621 &  -0.21122& 1\\
0.08571 &  0.11377 & -0.33905 & -1 &    0.02427 &  0.09378 & -0.20259 & 1\\
0.02248 & 0.24908 &  0.02209 & -1 &      0.08327 &  0.41050 &  0.14460 & 1\\
0.11506 &  -0.42290 &  -0.28321 & -1 &  0.17642 & -0.46325 & -0.25470 & 1\\
0.10066 &  0.47611  & -0.16828  & -1 &    0.17858 & -0.43237 &  0.16777 & 1\\
0.16194 & -0.08246 &  0.10723 & -1 &     0.14978 & -0.17273 & -0.32600 & 1\\
0.16014 & -0.15698 &  0.25501 & -1 &    0.18923 & -0.11888 &  0.24108 & 1\\
0.14219 & -0.14941  & 0.41892 & -1 &      0.16823 &  0.01108 & -0.47117 & 1\\
0.15487 &  -0.04512 & -0.40350 & -1&     0.02898 & -0.06648  &-0.22973& 1\\
 0.16316 & -0.07387  & 0.36614  &  -1&     0.13468 &  0.16391 & -0.43693 & 1\\
0.10994 & 0.09552  & -0.37405  &  -1&    0.16818 &  0.13211 &  0.30071 & 1\\
0.13900 &  0.15010 &  0.42616 & -1&      0.17777 &  0.06257 & -0.48490 & 1\\
0.08245 &  0.05951 & -0.19423 & -1 &      0.15628 &  0.30849 &  -0.08843 & 1\\
0.21808 &  0.26116 &  0.43648  & -1&        0.17012 &  0.26510 &   0.16623& 1\\\hline
\end{tabular}
}
\end{table}


\begin{table}[]
\centering
\caption{The list of positions of $28$ pairs of Weyl nodes with the Chiralities $+1$ and $-1$ at the scf Fermi level.}
\label{tab:weyllist1}
 {

\begin{tabular} {|c|c|c|c|c|c|c|c|}
\hline
k$_{x}$  & k$_{y}$ & k$_{z}$ & $\chi$ & k$_{x}$  & k$_{y}$ & k$_{z}$ & $\chi$  \\\hline
0.13173 &  0.36804 & -0.05039 & -1 &      0.16812 &  0.44582  & -0.12476 & 1 \\
0.13331 &  0.42331 &  -0.05322 & -1 &      0.17057 &  0.45672 & -0.00488  & 1\\
0.23242 &  0.49970 & -0.21677& -1 &    0.23282 & -0.49667 &   0.20653 & 1\\
0.16954 & -0.43711 & -0.01072 & -1&    0.21885 & -0.21880 & -0.12749 & 1 \\
0.24229 & -0.49718 &  0.15926 & -1&   0.23638 & -0.20900 &  0.01453 & 1\\
0.28717 & -0.17543  & 0.31547& -1&   0.20389 &  0.07254 &  -0.49298 & 1\\
0.23917 & -0.21492 &  0.46654& -1 &  0.23149 &  0.02521 & -0.04653 & 1\\
0.21997 &  -0.07856 & -0.49609 & -1&   0.18190 &  -0.12075 &  0.44811 & 1\\
0.17826 &  0.01345 & -0.48495 &  -1&   0.20715 &  0.00068  & 0.47723 & 1\\
0.24731 & -0.00187 & -0.03217 &  -1&    0.28466 &  0.16094 &  0.10693 & 1\\
0.23582 &  0.00945 &  0.20774 &  -1&   0.26154 &  0.36784  & 0.33681 & 1\\
0.23795 &  0.20373 &  0.03035 &   -1& 0.34188 & -0.11240  & -0.16960 & 1 \\
0.26376 &  0.35434 &  0.32361 &  -1&   0.35642 & -0.10525 & -0.05041  & 1\\
0.22869 &  0.39951 & 0.29596  &  -1&     0.29532 & -0.11459  & 0.15467  & 1 \\
0.20414 &  0.49996 & -0.03789 &  -1&     0.38172 & -0.18687 &  0.30177  & 1\\
0.30183 & -0.37359 &  0.13740 &  -1&     0.37873 & -0.02182 & -0.32114  & 1\\
0.31685 & -0.16001 &  0.24500 &  -1&      0.28847 &  0.09103  & 0.15510  & 1\\
0.31122 & -0.14148 & -0.08534 &  -1&      0.23945  & 0.21576 & -0.48067  & 1\\
0.33985 & -0.10028 &  0.07277 &   -1&     0.32639  & 0.40820 &  0.22745  & 1\\
0.38950 & -0.02893 & 0.30918&     -1&     0.47498  & -0.21738 & -0.00342 & 1\\
0.38661 &  0.09137 & -0.30382 &  -1&     0.47260 & -0.08904 & -0.02469  & 1\\
0.29694 &  0.11203 & -0.16228 &   -1&    0.40987 & -0.12614  & 0.06900  & 1\\
0.40424 & -0.16188 &  0.29095&    -1&       0.38384 &  0.12878 & -0.29565  & 1\\
0.46060 & -0.14710 &  0.03758 &  -1&       0.40021 & 0.10096  & 0.29554   & 1\\
0.46045 &  0.10352 &  0.18358  &  -1 &  0.39859 &  0.12551 & -0.07809 & 1 \\
0.46620 &  0.18266 & -0.24773  &  -1 &  0.48298 &  0.21020 &  0.00499 & 1 \\
0.43470 &  0.16997 &  0.27489  &  -1 &  0.40154 &  0.17748 &  0.30918 & 1 \\
0.42519 &  0.38493 &  0.36176  &  -1 &  0.41169 &  0.44436 &  0.41770 & 1 \\\hline
\end{tabular}
}
\end{table}
